\documentclass[final,3p,times]{elsarticle}

\usepackage{amsmath}
\usepackage{algorithm}
\usepackage{algpseudocode}
\usepackage[utf8]{inputenc}
\usepackage{stackengine}
\usepackage{textcomp}
\usepackage{xcolor}
\usepackage[skip=2pt,font=scriptsize]{caption} 
\usepackage{subcaption}
\usepackage{multirow}
\newcommand\rd{\mathrm d}
\usepackage{geometry,mathtools}
\usepackage{soul}

\biboptions{numbers,sort&compress}

\newtheorem{remark}{Remark}
\allowdisplaybreaks

\usepackage{setspace}
\makeatletter
\def\ps@pprintTitle{%
  \let\@oddhead\@empty
  \let\@evenhead\@empty
  \let\@oddfoot\@empty
  \let\@evenfoot\@oddfoot
}
\makeatother

\begin{document}

\begin{frontmatter}

\title{BattX: An Equivalent Circuit Model for Lithium-Ion Batteries Over Broad Current Ranges}

\author[label1]{Nikhil~Biju} \author[label1]{Huazhen~Fang\corref{cor1}}
\address[label1]{Department of Mechanical Engineering, University of Kansas, Lawrence, KS 66045, USA}
 \cortext[cor1]{Corresponding author. E-mail address: fang@ku.edu.}

\begin{abstract}
Advanced battery management is to  lithium-ion battery systems as   the brain is to the human  body. Its performance rests on the use of  battery models that are  both fast and accurate. However,   mainstream equivalent circuit models and electrochemical models have yet to meet this need well, due to   struggle with either  predictive accuracy or computational complexity. This problem has acquired   urgency as some emerging battery applications running across broad current ranges, e.g.,  electric vertical take-off and landing aircraft, can hardly find usable models from the literature. Motivated to address the problem, we  develop an innovative model in this study. Called {\em BattX}, the model  is  an equivalent circuit model  but  draws comparisons to a single particle model with electrolyte and thermal dynamics, thus combining their respective merits to be computationally efficient, accurate, and physically interpretable. The model design pivots on leveraging multiple circuits to approximate major electrochemical and physical processes in charging/discharging.  
Given the model, we develop a multipronged approach to design experiments and identify its  parameters   in groups from experimental data. Experimental validation proves that the BattX model is capable of accurate voltage prediction  for charging/discharging across low to high C-rates.  
\end{abstract}

\end{frontmatter}


\section{Introduction}

Lithium-ion batteries (LiBs)  are a key power source for  consumer electronics, electrified transportation,   smart grids, and  renewable energy. Compared with alternative battery electrochemistries, they provide a set of outstanding features, including high energy/power density, high nominal voltage, no memory effect, low self-discharge rates,
and long cycle life~\citep{Plett:BMSVol1:2015, Chaturvedi:IEEE:2010,Wang:CSM:2017}. Recent technological advances  have further improved their power performance and cost efficiency for a wider  application spectrum. High-quality dynamic models  are foundational to monitoring and control of LiBs for guaranteed  operational  safety and performance. While the growing research has led to a variety of useful models, 
the literature still lacks  fast and accurate models for applications involving charging/discharging from low to high current rates. To fill this gap, we propose a first-of-its-kind equivalent circuit model named   {\em BattX}  and demonstrate its predictive fidelity over broad C-rate ranges. 


{\em Literature Review.} Research on LiB dynamic modeling has flourished in the past decades to produce  a vast literature. The mainstream models generally fall into two categories: electrochemical models, and equivalent circuit models (ECMs). Electrochemical models explicitly describe electrochemical reactions,   transport of lithium ions, and distribution of charge and potential inside a LiB cell. Depending on the need for accuracy, they exist in diverse scales, from atomic/molecular to species level, and in different dimensions, from 1D to 3D and beyond, and often coupled with different physical processes, e.g., thermodynamics and stress/strain~\cite{Ramadesigan:JES:2012}. Generally, electrochemical models present high mechanistic fidelity as well as high computational complexity. Battery management researchers hence must  selectively focus  on those that offer desirable  accuracy-computation trade-off, due to practical demands for fast computation. A favorable choice  is the  pseudo-2D    Doyle-Fuller-Newman (DFN) model, which describes  the diffusion  of lithium ions and charge transfer across the electrodes, electrolyte, and separator of a sandwich cell~\cite{Doyle:JES:1993}. The search for more efficient models has led to the single particle model (SPM), which represents each electrode by a single spherical particle  and neglects the electrolyte dynamics~\cite{Ning:JES:2004}. The simplification enhances computational efficiency to a great extent but also limits the SPM model to low to moderate C-rates (around or less  than 1 C). Subsequent studies have emerged to expand  the SPM model by adding characterization of a cell's thermal behavior~\cite{Guo:JES:2011, Tanim:Energy:2015}, electrolyte dynamics~\cite{Rahimian:JPS:2013, Moura:TCST:2016, Han:JPS:2015, Marquis:JES:2019, Li:JES:2017}, stress buildup~\cite{Li:JES:2017}, or degradation~\cite{Li:AE:2018}, to elevate its  prediction capability. The literature has also presented a few computational methods to speed up simulation of the SPM model or its improved versions~\cite{Gopalakrishnan:TCST:2022,Saccani:JES:2022}. 

ECMs represent another important pathway to  modeling  LiBs.  They are circuit analogs composed of electrical components to simulate a cell's dynamic behavior, capture   phenomena in charging/discharging,  and track   state-of-charge (SoC) and power capability. With simple  structures, they are accessible to interpretation, easy to calibrate, and scalable to large LiB systems composed of many cells. Also, they are governed by low-order ordinary differential equations, thus allowing for very  fast computation. These benefits combine to make them popular candidates  for real-world battery management systems with limited computing resources~\cite{Farmann:AE:2018}. A basic ECM, called the Rint model,   cascades an open-circuit voltage (OCV) source with an internal resistor, in which the voltage source is SoC-dependent~\cite{He:Energies:2011}. One can add to the Rint model  a set of serially connected RC
pairs to  describe  the transient behavior in a cell's voltage response, leading to the so-called Thevenin's model~\cite{Mousavi:RSER:2014,Plett:Artech:2015}.  Depending on the number of RC pairs used,  one can set the model to capture transients at multiple time scales~\cite{Tian:JES:2020}. The literature has presented a few approaches to modify the Thevenin's model for better accuracy. For example, the study in~\cite{Plett:Artech:2015,Wang:CSM:2017}  incorporates   hysteresis in charging/discharging; in~\cite{Lee:TII:2018, Chen:TEC:2006, Hu:JPS:2012,Weng:JPS:2014},  different circuit parameters (e.g., the internal resistance) are made dependent on the SoC, temperature,  or    current loads, and  the OCV is  parameterized using different function forms for higher fitting accuracy. Even though phenomenological ECMs and electrochemical models were largely two disparate threads  of research,  a growing number of studies have explored to develop ECMs drawing upon  electrochemical modeling.   
The work in~\cite{Tian:IEEE:2020,Tian:IECON:2018} proposes the nonlinear double capacitor model  to approximate the ion diffusion in the electrodes of a cell and characterizes the nonlinear voltage behavior simultaneously. This model  is interpretable as a reduced-order version of the SPM, and it is  further supplemented   in~\cite{Movahedi:Mechatronics:2022}  with a data-based voltage hysteresis model to attain better accuracy. The study  in~\cite{Li:JPS:2021}  derives an ECM using circuit elements   to characterize charge transfer and diffusion potentials; the derivation also helps explain some conventional ECMs from an electrochemical perspective. In~\cite{Fan:AE:2022}, an ECM is coupled with diffusion dynamics to attain higher prediction accuracy. 
 
Structural simplicity  underlies the wide use of ECMs in battery management, but  also restricts their   accuracy. Most of today's ECMs  are   accurate enough for only low C-rates, and recent progress   has led to ECMs that are provably suitable for  about 1 C~\cite{Tian:IEEE:2020,Tian:IECON:2018}. However, the literature still faces an absence of ECMs  capable of predicting a cell's voltage  behavior from  low to high C-rate ranges. This gap will pose potential barriers for emerging  battery-powered applications that  must operate  across wide current ranges. One example is electric vertical take-off and landing (eVTOL), which   requires discharging of up to 5 C in the take-off and landing phases and necessitates precise   models to  fulfill  high-stakes safety requirements~\cite{Sripad:arXiv:2020}. 

{\em Statement of Contributions.} To address the above challenge, we   develop a new  ECM that promises to predict over  broad current ranges. Our work takes inspirations from electrochemical modeling to   design  and conjoin circuits to   simulate a LiB cell's electrode, electrolyte, and thermal dynamics as well as their effects on the terminal voltage. Here, we select the SPM with electrolyte and thermal dynamics (SPMeT) as the reference  benchmark. The  obtained ECM, called BattX, hence is physically comparable and represents a reduced-order analog to  the SPMeT model. By design, the BattX model comprehensively  accounts  for the aforementioned  different types of dynamics that have a phenomenologically  appreciable impact  at high C-rates. This endows it with not only excellent prediction capability, but also considerable  physical  fidelity and interpretability.  Desirably, the model still retains relatively compact structures to present high computational efficiency, carrying a potential to facilitate embedded  battery management systems.  
To sum up, this paper delivers  the following specific contributions. 

\begin{itemize}
\item  We propose the principled    design of the BattX model and further elucidate the underlying rationale by showing its connections with the SPMeT model in detail.  

\item We develop a multipronged parameter identification approach to extract   the parameters of the BattX model from measurement data made on LiBs. The approach will make the model readily available in practice. 

\item We provide  experimental evaluation results to validate the effectiveness and accuracy of the BattX model. The experiments involve charging/discharging at high C-rates and consider operation profiles of eVOTL as a case study. 
\end{itemize}

{\em Organization.} The rest of the paper is organized as follows. Section \ref{sec:MD} presents the  BattX model design as a whole. Section \ref{sec:Deriv} proceeds to elucidate on the model's correspondence to the SPMeT model. Section \ref{sec:ParamID} develops the parameter identification pipeline of the model. Section \ref{Sec:Validation}   evaluates    the model using experimental data. Finally, Section \ref{Sec:Conclusion} offers concluding remarks. 

\section{The BattX Model}\label{sec:MD}

This section presents the structure and governing equations of the BattX model. We will provide the detailed rationale for the model  design subsequently in Section~\ref{sec:Deriv}.

\begin{figure}[t]
\centering
\includegraphics[width = 0.85 \textwidth]{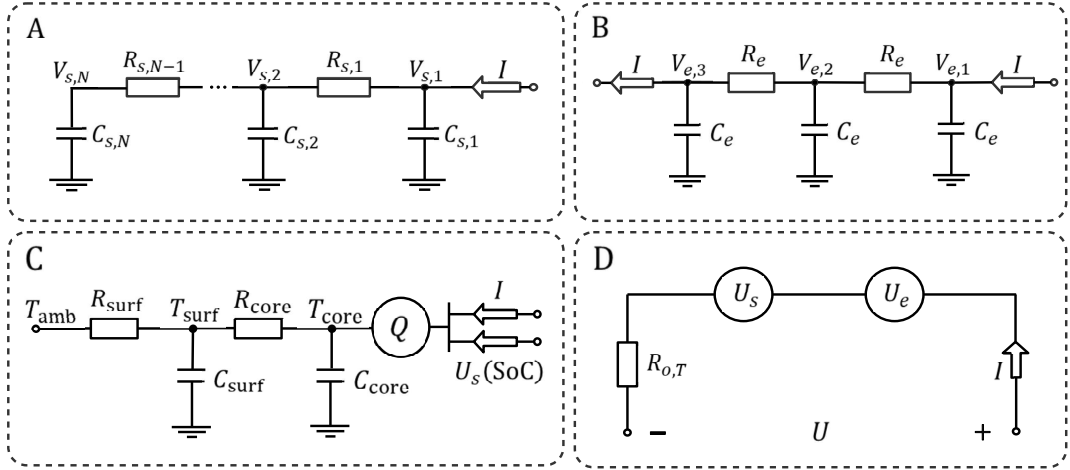}
\caption{The BattX model comprising: sub-circuit A to simulate the  lithium-ion diffusion in the electrode phase; sub-circuit B to simulate the lithium-ion diffusion in the electrolyte phase; sub-circuit C to simulate heat conduction and convection; and sub-circuit D to simulate the   terminal voltage.}
\label{XF}
\end{figure}

 At the core, the BattX model attempts to characterize the multiple major dynamic processes innate to a LiB cell in order to capture the cell's behavior from low to high current rates. This is akin to electrochemical modeling to a certain extent, but a main difference is that the BattX model leverages  circuit analogs to simulate the  processes.  Fig.~\ref{XF} shows the overarching structure of the model. As is seen, it consists of four coupled sub-circuits, which  are   labeled as A to D. These sub-circuits are designed to  approximate  the  cell's electrode-phase  diffusion, electrolyte-phase  diffusion, thermal evolution, and voltage response, respectively. 


To begin with, sub-circuit A uses a chain of resistors and capacitors to approximate the lithium-ion  diffusion  in the electrode phase. Its governing equations are
 \begin{subequations}\label{Sub-circuit-A-eqns}
\begin{align}\label{V1}
\dot{V}_{s,1}(t) &= \frac{V_{s,2}(t) - V_{s,1}(t)}{C_{s,1}R_{s,1}} + \frac{I(t)}{C_{s,1}},\\  \label{Vi}
\dot{V}_{s,i}(t) &= \frac{V_{s,i-1}(t) - V_{s,i}(t)}{C_{s,i}R_{s,i-1}} + \frac{V_{s,i+1}(t) - V_{s,i}(t)}{C_{s,i}R_{s,i}}, \; \;  i = 2, \ldots, N-1\\
 \label{VN}
\dot{V}_{s,N}(t) &= \frac{V_{s,N-1}(t) - V_{s,N}(t)}{C_{s,N}R_{s,N-1}},
\end{align}
\end{subequations} 
where $I$ is the applied current, with $I>0$ for charging and $I<0$ for discharging, $V_{s,j}$ for $j=1,\ldots, N$ are the voltages  across the individual capacitors $C_{s,j}$,   $R_{s,j}$ are the resistors that the current must flow through, and the subscript $s$ refers to the solid phase. We set $0\leq V_{s,j} \leq 1$ for the purpose of normalization and then define the SoC as the percentage ratio of the currently available charge over the total charge capacity, which is 
\begin{align*}
\mathrm{SoC} = \frac{\sum_{j=1}^{N} C_{s,j} V_{s,j} }{\sum_{j=1}^{N} C_{s,j} } \times 100\%. 
\end{align*} 
That is, $\mathrm{SoC} = 100\%$ when $V_{s,j} = 1$ for all $j$, and $\mathrm{SoC} = 0$ when $V_{s,j} = 0$ for all $j$. 
A brief interpretation of sub-circuit A  is  as  follows, with more details to be shown in   Section~\ref{sec:Deriv}. Overall, the charge transfer between the capacitors in the circuit mimics the diffusion of lithium ion in the solid phase or electrode. Then,  $V_{s,j}$ for $j=1,\ldots,N$      correspond  to the lithium-ion concentrations at $N$ different locations, from the surface to the center,  that spread along the radius of an electrode sphere; $C_{s,j}$ for $j=1,\ldots,N$ are analogous  to  the volumes of the subdomains if one subdivides the electrode sphere  at these discrete locations; $R_{s,j}$ for $j=1,\ldots,N-1$  resist the charge transfer or equivalently, the solid-phase diffusion in the SPMeT model, and   are hence inversely proportional to the diffusivity. 


Along similar lines to sub-circuit A, sub-circuit B uses a resistor-capacitor chain   to approximate the lithium-ion diffusion in the electrolyte. Its dynamics is governed by 
\begin{subequations}\label{Sub-circuit-B-eqns}
\begin{align}\label{Anode Elec}
\dot{V}_{e,1}(t) &=   \frac{V_{e,2}(t)-V_{e,1}(t)}{C_eR_e}+ \frac{I(t)}{C_e},\\\label{Sep Elec}
\dot{V}_{e,2}(t) &= \frac{V_{e,1}(t)-2V_{e,2}(t)+V_{e,3}(t)}{C_eR_e},   \\\label{Cath Elec}
\dot{V}_{e,3}(t) &= \frac{V_{e,2}(t)-V_{e,3}(t)}{C_eR_e} - \frac{I(t)}{C_e},
\end{align}
\end{subequations}
wehre the notations in above have similar meanings as in~\eqref{Sub-circuit-A-eqns},  and  the subscript $e$ refers to the electrolyte. We let $0 \leq  V_{e,j} \leq 1$ for  $j=1,2,3$ as in the case of $V_{s,j}$, and further assume that  $ V_{e,j} = 0.5$ for $j=1,2,3$ when the cell is at equilibrium. One can interpret sub-circuit B as analogous to the  one-dimensional electrolyte-phase diffusion 
 that is   discretized along the spatial coordinate. In particular, $V_{e,j}$ for $j=1,2,3$ can be associated with the lithium-ion concentrations at the   locations of the anode, separator, and cathode, and $R_e$ embodies resistance to the diffusion. The spatial discretization is assumed to be   uniform, thus leading to the same values of $R_e$ and $C_e$ for each region as shown in~\eqref{Sub-circuit-B-eqns}.  


Sub-circuit C is a lumped circuit model for the thermal dynamics, with the design inspired by~\cite{Lin:JPS:2014}. Here, we consider the cell to be a cylindrical one  without loss of generality  and concentrate its  spatial
dimensions into two singular points that represent the surface and core, respectively. This simplification allows to describe the evolution of    the temperatures at these two points, $T_{\mathrm{surf}}$ and $T_{\mathrm{core}}$,  by 
\begin{subequations}\label{Lumped-Thermal-Model}
\begin{align}\label{Core-Temp}
\dot{T}_\mathrm{core}(t) &= \frac{Q(t)}{C_\mathrm{core}} + \frac{T_\mathrm{surf}(t) - T_\mathrm{core}(t)}{R_\mathrm{core} C_\mathrm{core}}  , \\
\label{Surface-Temp}
\dot{T}_\mathrm{surf}(t) &= \frac{T_{\mathrm{amb}}(t) - T_\mathrm{surf}(t)}{R_\mathrm{surf}C_\mathrm{surf}} - \frac{T_\mathrm{surf}(t) - T_\mathrm{core}(t)}{R_\mathrm{core}C_\mathrm{surf}} ,
\end{align}
\end{subequations}
where $T_{\mathrm{amb}}$ is the ambient temperature,   $C_\mathrm{surf/core}$ and $R_\mathrm{surf/core}$ represent the thermal capacitance and resistance at the surface and core, respectively, and $Q$ is the internal heat generation rate accompanying electrochemical reactions   inside the cell during charging/discharging. From a heat transfer perspective,~\eqref{Core-Temp} approximately describes the heat conduction between the cell's surface and core, and~\eqref{Surface-Temp} grasps the convection between the surface and the ambient environment.  Further, $Q$ is characterized as 
\begin{align}\label{Qdot Model}
Q = -I \left[ U_s \left(\mathrm{SOC}\right) - U_s(V_{s,1}) - R_{o,T}I \right], 
\end{align} 
 where $U_s(\cdot)$ is the nonlinear  OCV function,  $V_{s,1}$ is defined in sub-circuit A,  and $R_{o,T}$ is the internal resistance.  


Finally, sub-circuit D summarizes the effects of the solid-phase and  electrolyte-phase dynamics on the terminal voltage. It contains two voltage sources, $U_s$ and $U_e$, in series with an internal resistance $R_{o,T}$. The terminal   $U$  is given by 
\begin{align}\label{Terminal}
 U = U_s(V_{s,1}(t)) + U_e(V_{e,1}(t),V_{e,N_e}(t)) + R_{o,T}I(t).
\end{align}
Here, $U_s$ simulates the solid-phase OCV. As the SPMeT model mandates that  the open-circuit potential of solid material relies on the lithium-ion concentration at the surface of the electrode,  $U_s$ should come  as a function of $V_{s,1}$, and its exact  form  will depend  on the cell. For the cell used in our experiments in Sections~\ref{sec:ParamID}-\ref{Sec:Validation},   we find the following parameterization of $U_s$ suitable:
\begin{align*}  
U_s(V_{s,1}) =   h_{1}\left(V_{s,1} \right) \cdot H\left(0.9 - V_{s,1}\right) + h_{2}\left(V_{s,1}\right) \cdot H \left(V_{s,1}-0.9\right),
\end{align*}
where $H(\cdot)$ is the Heaviside step function, $h_1(V_{s,1})$ captures the behavior when $V_{s,1} \leq 0.9$ as
\begin{align*}
h_{1}(V_{s,1}) &= \alpha_0 + \alpha_1 \frac{1}{1+\mathrm{exp}(\alpha_2(V_{s,1}(t) - \alpha_3))} +  
\alpha_4 \frac{1}{1+\mathrm{exp}(\alpha_5(V_{s,1} - \alpha_6))} + \alpha_7 \frac{1}{1+\mathrm{exp}(\alpha_8 (V_{s,1} - \alpha_9))} + \\ 
&\quad \alpha_{10} \frac{1}{1+\mathrm{exp}(\alpha_{11}V_{s,1}(t))} + \alpha_{12}V_{s,1}(t) , 
\end{align*}
and $h_2(V_{s,1})$ is for when $0.9<V_{s,1}\leq 1$ with
\begin{align*}
h_{2}(V_{s,1}) = \alpha_{13} \mathrm{exp}(\alpha_{14}V_{s,1}) + \alpha_{15} \mathrm{exp}(\alpha_{16} V_{s,1}). 
\end{align*}
Here, $\alpha_i$ for $i=0,\ldots,15$ are constant coefficients. 
Next, we need to determine the form of $U_e$. In the SPMeT model, the electrolyte potential depends on the electrolyte concentration at the anode and cathode. We hence  make   $U_e$ as a function of   $V_{e,1}$ and $V_{e,N_e}$ and  express it  as 
\begin{align}\label{Elec-Potential}
U_{e}(t)=\beta_1\left(\mathrm{ln}\left(\frac{ V_{e,1}(t)+\beta_2}{V_{e,3}(t)+\beta_2}\right)\right),
\end{align}
where $\beta_i$ for $i= 1,2$ are constant coefficients. 
As the last element of the model,  $R_{o,T}$  is  not a constant and instead depends on  $\mathrm{SoC}$ and  $T_{\mathrm{core}}$.  It is  given by
\begin{align}\label{Ro}
R_{o.T} = R_o(\mathrm{SoC}) \cdot  \mathrm{exp}\left(\kappa_1 \left(\frac{1}{T_{\mathrm{core}}} - \frac{1}{T_{\mathrm{amb}}}\right)\right),
\end{align}
where $\kappa_1$ is a constant coefficient. 
In above, the first term  $R_o(\mathrm{SoC})$ captures the dependence of  $R_{o,T}$  on $V_{s,1}$  and takes the form
\begin{align}
R_o(\mathrm{SoC}) =  \gamma_1 + \gamma_2 \cdot \mathrm{exp}\left(-\gamma_3 \mathrm{SoC}\right),
\end{align} 
where $\gamma_i$ for $i=1,2,3$ are coefficients, 
 and the second term shows the temperature dependence due to  the Arrhenius law. Similarly, an Arrhenius relationship can be used to capture the relationship between the electrode-phase diffusion constant and temperature:
\begin{align}\label{Rs1T}
R_{s,1,T} = R_{s,1} \cdot  \mathrm{exp}\left(\kappa_2 \left(\frac{1}{T_{\mathrm{core}}} - \frac{1}{T_{\mathrm{amb}}}\right)\right),
\end{align}

Putting together all the above equations, we will obtain a complete description of the BattX model.  
This model is the first ECM  that can predict over broad current ranges, due to the integration of the circuits approximating   the electrode, electrolyte, and thermal dynamics into a whole. 
The model design also leads to profound comparability with electrochemical modeling, especially the SPMeT, which will be revealed further in the next section. We will address the identification  of the model parameters in Section~\ref{sec:ParamID}.

\section{Rationale for the BattX Model Design}\label{sec:Deriv}

In this section, we will use the SPMeT model as a benchmark to explain the rationale for the design of the BattX model.   We will show that the SPMeT model,  if appropriately discretized,  will reduce  to a structure that is approximately equivalent to the proposed circuit analogs of the BattX model.   Our    main  references about the SPMeT model include~\citep{Mehta:EA:2021,Moura:IEEE:2016}.  We will focus  on expounding sub-circuits A, B,  and D, with the sub-circuit C-based lumped thermal model    well addressed in~\cite{Lin:JPS:2014}. 

\subsection{Connection between Sub-circuit A and SPMeT}\label{SubCircuitA_SPMeT}

The SPMeT model characteristically couples the SPM model with the electrolyte and thermal dynamics. What it inherits  from  the SPM model is the representation of the electrodes as two spherical particles. The  diffusion of lithium-ions in each  particle follows Fick's second law in spherical coordinates~\cite{Guo:JES:2011,Tanim:Energy:2015}:
\begin{align}\label{Solid-phase-diffusion}
\frac{\partial c_{s,j}(r,t)}{\partial t} = \frac{D_{s,j}}{r^2}\frac{\partial}{\partial r}\left(r^2\frac{\partial c_{s,j}(r,t)}{\partial r}\right),
\end{align}
where $c_{s,j}$ is  the solid-phase (electrode) lithium-ion concentration,   $D_s$ is the constant diffusion coefficient, and $r$ is the radial coordinate. The subscript $j \in \{n, p\}$, where  $n$ and $p$ refer to the anode (negative) and cathode (positive), respectively. The boundary conditions for~\eqref{Solid-phase-diffusion} are 
\begin{align*}
\left.\frac{\rd c_{s,j}}{\rd r}\right|_{r=0} = 0, \; \;  \left.\frac{\rd c_{s,j}}{\rd r}\right|_{r=R_j} = -\frac{J_j}{D_{s,j}},
\end{align*}
where $R$ is the radius of a particle. The molar flux $J$     at the electrode/electrolyte interface is given by
\begin{align*}
J_p(t) = \frac{i(t)}{FS_p}, \; \; J_n(t) = -\frac{i(t)}{FS_n}
\end{align*}
 where $i$ is the applied current density, with $i > 0$ for charging and $i < 0$ for discharging, $S$ is the surface area of a particle, and $F$ is Faraday's constant. 

\begin{figure}[t]
\centering
\includegraphics[width = 0.4\textwidth]{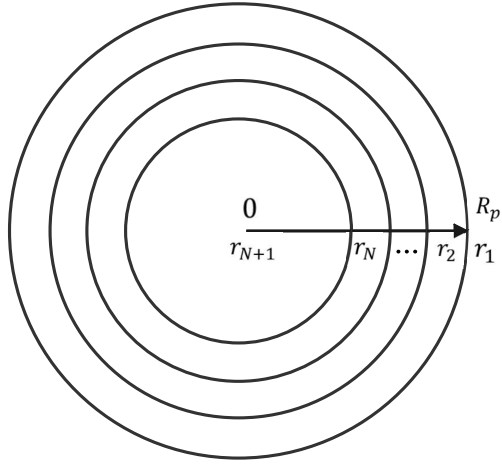}
\caption{Spherical discretization of an electrode particle. }
\label{Discretized Sphere}
\end{figure}

Next, let us reduce the PDE in~\eqref{Solid-phase-diffusion}  into  a system of ODE equations using a finite volume method~\cite{Fang:TCST:2017,Xu:JES:2016}. The subscript $j \in {n,p}$ will be dropped in sequel    without causing confusion. First, we   subdivide the particle into a set of continuous finite volumes at discrete locations $r_1 =R > r_2 > \ldots > r_{N}> r_{N+1}=0$  that spreads inward from the surface to the center, as show in Fig.~\ref{Discretized Sphere}.    The   lithium-ion amount within the $i$th finite volume is given by
\begin{align*}
Q_i(t)  = \int_{r_{i+1}}^{r_i} c_s(r,t)  d V  = \int_{r_{i+1}}^{r_i} c_s(r,t) \cdot 4 \pi r^2 dr,
\end{align*}
for $i=1,\ldots,N$. 
Then, using~\eqref{Solid-phase-diffusion}, we have
\begin{align}\label{Q_dynamics}
\dot Q_i(t)  = \int_{r_{i+1}}^{r_i} \dot c_s(r,t) \cdot 4 \pi r^2 dr =  \int_{r_{i+1}}^{r_i}    d \left( 4 \pi D_s  r^2 \frac{\partial c_s(r,t)}{ \partial r} \right)  =  4 \pi D_s  r_i^2 \left. \frac{\partial c_s(r,t)}{ \partial r} \right|_{r_i} - 4 \pi D_s  r_{i+1}^2 \left. \frac{\partial c_s(r,t)}{ \partial r} \right|_{r_{i+1}}. 
\end{align}
To proceed, we replace $c_s(r,t)$ by the average lithium-ion concentration within the $i$th finite volume, $\bar c_s(r_i, t)$:
\begin{align}\label{average-concentration-def}
\bar c_s(r_i, t) = \frac{Q_i(t)}{\Delta V_i},
\end{align}
where $\Delta V_i = 4 \pi (r_i^3 - r_{i+1}^3)/3$. From~\eqref{Q_dynamics}-\eqref{average-concentration-def}, it follows that
\begin{align*}
\dot {\bar c}_s (r,t) = \frac{ 4 \pi D_s  r_i^2}{\Delta V_i} \left. \frac{\partial c_s(r,t)}{ \partial r} \right|_{r_i} -  \frac{4 \pi D_s  r_{i+1}^2 }{\Delta V_i} \left. \frac{\partial c_s(r,t)}{ \partial r} \right|_{r_{i+1}}. 
\end{align*}
Then, we approximate the concentration gradient along the radial coordinate as 
\begin{align*}
\left. \frac{\partial c_s(r,t)}{\partial r} \right|_{r_i} = \frac{\bar c(r_{i-1}, t)  -\bar c(r_{i }, t) }{\Delta r_i}, 
\end{align*}
where $\Delta r_i = (r_{i-1} - r_{i+1})/2$. 
Given the boundary conditions, we further have
\begin{subequations}\label{solid-phase-diffusion-ODEs}
\begin{align}
\dot {\bar c}_s(r_1, t)  &= - \frac{4 \pi D_s r_2^2}{\Delta V_1 \Delta r_2} \left( \bar c_s(r_1, t)  -\bar c_s( r_2 , t)  \right) +  \frac{4 \pi   r_1^2}{\Delta V_1 FS}  i(t),  \\
\dot {\bar c}_s(r_i, t)  &= \frac{4 \pi D_s r_i^2}{\Delta V_i \Delta r_i} \left( \bar c_s(r_{i-1}, t)  -\bar c_s(r_{i }, t)  \right)  - \frac{4 \pi D_s r_{i+1}^2}{\Delta V_i \Delta r_{i+1} } \left( \bar c_s(r_i, t)  -\bar c_s( r_{i+1}  , t)  \right) , \;\; i=2,\ldots, N-1,\\
\dot {\bar c}_s(r_{N}, t)  &= \frac{4 \pi D_s r_{N}^2}{\Delta V_{N}  \Delta r_{N}} \left( \bar c_s(r_{N-1}, t)  -\bar c_s(r_{N}, t)  \right). 
\end{align} 
\end{subequations}

The above ODEs show the spatially discretized solid-phase diffusion. Note that they share the same structure with~\eqref{Sub-circuit-A-eqns}. A closer inspection of~\eqref{Sub-circuit-A-eqns} and~\eqref{solid-phase-diffusion-ODEs} suggests: 1) $V_s$ is a mirror of $\bar c_s(r,t)$, and its distribution   reflects the distribution of lithium-ion concentrations inside an electrode particle; 2) $C_s$ is a mirror of $\Delta V$, associating the capacitance with the volume of a finite volume element within the particle; 3) $R_s$ roughly corresponds to $ \Delta r / (D_s \cdot 4 \pi r^2)$ to grasp the effect of $D_s$, $\Delta r$ and $r$ on the diffusion resistance at different   locations. This unveiled connection with the SPMeT model justifies the design of  sub-circuit A. 
 
\subsection{Connection between Sub-circuit B and   SPMeT} 
\label{Sub-circuit-B-SPMeT}

\begin{figure}[t]
\centering
\includegraphics[width = 80mm]{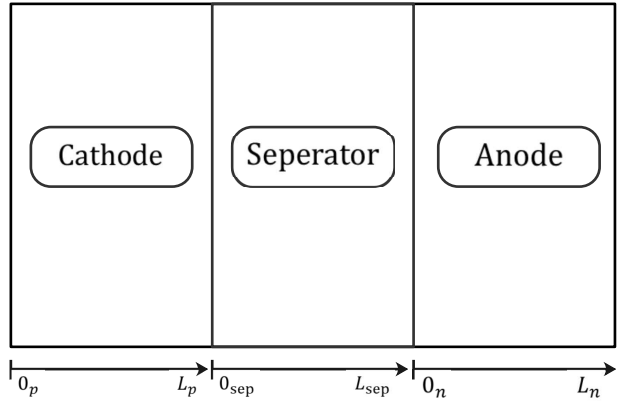}
\caption{Three regions immersed in the electrolyte.}
\label{fig:ElectDisc}
\end{figure}

The SPMeT model includes   one-dimensional   electrolyte diffusion, which also follows Fick's second law. The electrolyte diffusion is considered in the electrode and separator domains that are all immersed in the electrolyte.  Based on the coordinates in each domain as shown in Fig.~\ref{fig:ElectDisc}, the governing equations are
\begin{subequations}\label{Electrolyte-phase-diffusion}
\begin{align}\label{Cep}
\frac{\partial c_{e,p}(x,t)}{\partial t} &= D_e \frac{\partial^2 c_{e,p}(x,t)}{\partial x^2} + \frac{1 - t^0 _c}{\epsilon_{e,p}FL_p}i(t),\\\label{Cesep}
\frac{\partial c_{e,\mathrm{sep}}(x,t)}{\partial t} &= D_e \frac{\partial^2 c_{e,\mathrm{sep}}(x,t)}{\partial x^2},\\\label{Cen}
\frac{\partial c_{e,n}(x,t)}{\partial t} &= D_e \frac{\partial^2 c_{e,n}(x,t)}{\partial x^2} - \frac{1 - t^0 _c}{\epsilon_{e,n}FL_n}i(t),
\end{align}
\end{subequations}
where $c_{e,j}$ for $j \in \{n, p, \mathrm{sep} \}$ is the lithium-ion concentration in the electrolyte surrounding the anode, cathode and separator, $\epsilon_{e,j}$ is  the electrolyte volume fraction, $D_{e,j}$ is the electrolyte diffusion coefficient, and $t^0_c$ is the constant transference number.  We assume that $\epsilon_{e,j}$  and $D_{e,j}$  are  the same for any $j \in \{n, p, \mathrm{sep} \}$. 
The boundary conditions are given by
\begin{align*}
\frac{\partial c_{e,p}(0_p,t)}{\partial x} & = \frac{\partial c_{e,n}(L_n,t)}{\partial x} = 0, \\
\frac{\partial c_{e,p}(L_p,t)}{\partial x} & = \frac{\partial c_{e,\mathrm{sep}}(0_{\mathrm{sep}},t)}{\partial x}, \\
\frac{\partial c_{e,\mathrm{sep}}(L_{\mathrm{sep}},t)}{\partial x} & = \frac{\partial c_{e,n}(0_n,t)}{\partial x},  \\
c_{e}(L_p,t) & =c_{e}(0_\mathrm{sep},t), \\
c_{e}(L_{\mathrm{sep}},t) & = c_{e}(L_n,t). 
\end{align*} 

To convert~\eqref{Electrolyte-phase-diffusion} into ODEs, we   concentrate the electrodes and separator into singular points and further suppose   $L_p = L_n$ and   $L_\mathrm{sep}$ is negligible. The singular point that represents the electrodes are  located at the midpoint of each domain, and the average lithium-ion concentration is denoted as $\bar c_{e,j}$. Then, we apply the finite difference to~\eqref{Electrolyte-phase-diffusion} and obtain  
\begin{subequations}
\begin{align}
\dot {\bar c}_{e,p} (t) &= \frac{4 D_e}{L^2} \left (  \bar c_{e,\mathrm{sep} } (t) - {\bar c}_{e,p} (t) \right) + \frac{1-t^0_c}{\epsilon_{e }FL } i(t), \\
\dot {\bar c}_{e,\mathrm{sep}} (t) &= \frac{4 D_e}{L^2}\left( \bar c_{e,p} (t) -2 \bar c_{e,\mathrm{sep}} (t) + \bar c_{e,n} (t) \right), \\
\dot {\bar c}_{e, n} (t)  &= \frac{4 D_e}{L^2} \left (  \bar c_{e,\mathrm{sep}}(t) - {\bar c}_{e,n }(t) \right)  - \frac{1-t^0_c}{\epsilon_{e}FL } i(t). 
\end{align}
\end{subequations}

As is seen,~\eqref{Electrolyte-phase-diffusion} is structurally similar  to~\eqref{Sub-circuit-B-eqns}, and the similarity  lends to interpretation of~\eqref{Sub-circuit-B-eqns} through the  lens of electrochemical modeling. Specifically, we can associate $V_{e,1}$, $V_{e,2}$ and $V_{e,3}$ with ${\bar c}_{e,p}$,  ${\bar c}_{e,\mathrm{sep}}$ , and ${\bar c}_{e,n}$, respectively. Further, $C_e$ can be linked with the spatial lengths of the electrode domains, which decide the volume of the electrolyte, and $R_e$ comes as the inverse of $D_e$ to measure the resistance against electrolyte diffusion. 

\subsection{Connection between Sub-circuit D and SPMeT}

In the SPMeT model, the terminal voltage $V$ consists of four terms that represent   the solid-phase OCV, electrolyte-phase voltage,  overpotential, and voltage over the film resistance, respectively.  Then, coming back to sub-circuit D of the BattX model, $U_s$ mirrors the solid-phase OCV, $U_e$ corresponds to the electrolyte-phase voltage, and $R_{o,T}$ plays a role   to mainly capture the film resistance as well as the overpotential effect. Less trivially, we elaborate on the form of $U_e$ in~\eqref{Elec-Potential}. The electrolyte-phase voltage is given by
\begin{equation}\label{Phi e}
\phi_e(0_p,t) -  \phi_e(L_n,t) = \frac{L_p + 2L_{sep} + L_n}{2\bar{k}}i(t) + k_{\mathrm{conc}} \left( \mathrm{ln}c_e(0_p,t) - \mathrm{ln}c_e(L_n,t) \right),
\end{equation}
where $\phi_e$ is the electrolyte electric potential,  and $\bar{k}$ and $k_\mathrm{conc}$ are two coefficients that are related with electrolyte conductivity and molar activity. The first term in above is accounted by for $R_{o,T}$.  Following the discussion in Section~\ref{Sub-circuit-B-SPMeT}, we can approximate the second term as
\begin{align*}
k_{\mathrm{conc}} \left( \ln  \bar c_{e,p}(t) - \ln \bar c_{e,n}( t)\right).
\end{align*}
This form is found to bear equivalence to~\eqref{Elec-Potential}, when making   linear projections of $ \bar c_{e,p}(t) $ and  $\bar c_{e,n}(t)$ to $V_{e,1}$ and $V_{e,3}$, respectively.  

\section{Parameter Identification for the  BattX Model}\label{sec:ParamID}

In this section, we investigate how to determine   the parameters of the BattX model. To this end, we separate  the model's parameters into  different groups based on  the dynamic processes that they belong to or prominently influence. We then  design   experiments accordingly and use different current profiles to excite different dynamic processes  and obtain    voltage  or temperature  data suitable for the identification of the corresponding parameter groups. Finally, we extract the parameters from the data, group by group, through data fitting and some empirical  tuning.  

To begin with, we set up the following parameter groups   for the BattX model:
\begin{itemize}

\item $\Theta_{U_s} = \left\{  \alpha_i, i = 0, 1, \ldots, 16 \right\} $, which includes the parameters in $U_s$ in sub-circuit D;

\item $\Theta_{R_o} =  \left\{  \gamma_i, i=1,2,3 \right\}$, which includes the parameters in $R_o$  in sub-circuit D; 

\item $\Theta_{s}= \left\{ C_{s,i}, i =1,\ldots, N,  R_{s,j}, j = 1, \ldots,N-1 \right\}$, which includes the parameters of sub-circuit A;

\item $\Theta_{\mathrm{Th}}= \left\{  C_\mathrm{surf},  R_\mathrm{surf},  C_\mathrm{core}, R_\mathrm{core}  \right\}$, which includes the parameters in the lumped thermal model in  sub-circuit C; 

\item $\Theta_{e}= \left\{    C_e,  R_e,  \beta_1,  \beta_2  \right\}$, which includes the parameters in sub-circuit B and the parameters in $U_e$ in sub-circuit D;

\item $\Theta_{\mathrm{Arr}}= \left\{  \kappa_1,  \kappa_2 \right\}$, which includes the Arrhenius-law-related parameters;.

\end{itemize}
By grouping the parameters as above, we can design different current input profiles to stimulate different parts of the cell's dynamics so as to identify the parameters     group by group. This multi-pronged approach includes the following steps. 

{\em Step 1: Identification of $\Theta_{U_s}$.} Since $U_s$ represents the OCV source, we can capture  it by applying a trickle constant current with a magnitude of $1/30$ C to fully charge or discharge the cell. As the current is extremely small, sub-circuit A, which is an analog to the solid (electrode)-phase diffusion, is almost always at equilibrium, with $V_{s_i} = \mathrm{SoC}$ for $i=1,\ldots, N$ ($\mathrm{SoC}$ can be obtained via Coulomb counting); meanwhile,    sub-circuits B and C, $U_e$, and the voltage across $R_{o,T}$, are all negligible in this case. Hence,   $U \approx U_s$, and we can construct the following data fitting problem to identify $\Theta_{U_s}$:
\begin{align}\label{Data-Fitting-Theta-Us}
\hat \Theta_{U_s} = \arg \min_{\Theta_{U_s}} \sum_{t_k}  \left[ U(t_k)  - U_s \left ( \mathrm{SoC} (t_k) ; \Theta_{U_s} \right )  \right]^2,
\end{align}
where $k$ is the discrete time index in the experiment. 

{\em Step 2: Identification of $\Theta_{R_o}$.} $R_o$ is an integral part of the internal resistance $R_{o,T}$, and $R_o = R_{o,T}$ when $T = T_\mathrm{ref}$. To identify $\Theta_{R_o}$, we  apply a   0.5 C pulse current profile, which includes   long enough rest periods between two consecutive pulses to allow for sufficient voltage recovery, to discharge the cell from $100\%$ to $0\%$ of SoC when the ambient temperature is $T_\mathrm{ref}$. With discharging at 0.5 C, the cell will see only a negligible increase in its temperature, and $U_e\approx 0$. For the terminal voltage $U$, we will see a sharp drop  or jump at the beginning or end of every pulse, and this is almost solely due to the voltage change across $R_o$. Therefore, using the voltage jump, one can approximate  $R_o$ as
\begin{align}\label{Ro-Calculation}
\tilde R_o (t_*)  = \left|\frac{U(t_{*+1}) - U(t_*)}{I} \right|, 
\end{align}
where $t_*$ is the instant when a pulse stops. Further, the instantaneous SoC can be readily determined via Coulomb counting. Collecting $R_o$ for all $t_*$, we can formulate the following data fitting problem to estimate $\Theta_{R_o}$:
\begin{align}\label{Theta-Ro-Identify}
\hat  \Theta_{R_o}  =   \arg \min_{\Theta_{R_o}} \sum_{t_*} \left[ \tilde R_o (t_*)  - R_o (\Theta_{R_o}; t_*)\right]^2.
\end{align}

{\em Step 3: Identification of $\Theta_{s}$.} The number of parameters in $\Theta_s$ depends on $N$, and when $N$ is large,   $\Theta_s$ will be poorly identifiable to defy accurate estimation. To formulate a tractable   identification problem, we  assume that 
\begin{align}\label{Csi-Rsi-Relation}
C_{s,i} = \eta_i C_{s,1}, \; R_{s,j} = \sigma_j R_{s,1}, 
\end{align}
where $\eta_i$ and $\sigma_i$ for $i=1,\ldots N$ and $j=1, \dots N-1$ are pre-specified coefficients with $\eta_1 = \sigma_1 = 1$, and $\sum_{i=1}^N\eta_i C_{s,i}$ is the total capacity of the cell. This  allows us to consider only two parameters, i.e., $\Theta_s = \left\{ C_{s,1}, R_{s,1} \right \}$,  greatly facilitating the parameter estimation. The simplification is also reasonable---the difference among $C_{s,i}$ and $R_{s,j}$ can be viewed as a result of the selection of the discretization points as shown in~\eqref{solid-phase-diffusion-ODEs}, and one can specify   $\eta_i$ and $\sigma_j$ assuming that they result from a certain selection. The practical  selection of  $\eta_i$ and $\sigma_j$ can be through analysis of the discretization  shown in Section~\ref{SubCircuitA_SPMeT} and tuning.  Going forward, we apply a  0.5 C constant-current profile to discharge the cell from full to zero SoC.  In this   setting,  sub-circuit A is excited, but  the dynamics of sub-circuits B and C have no appreciable effects. That is, the cell's temperature     remains  almost the same, and $U_e \approx 0$.  We can conduct data fitting as below to find out $\Theta_s$:
\begin{align}\label{Thetas-Identify}
\hat \Theta_{s} = \arg \min_{\Theta_{s}} \sum_{t_k}  \left[ U(t_k) -  R_{o}\left( \hat \Theta_{R_o}; t_k \right)  I(t_k) - U_s \left (V_{s,1}\left({  \Theta_s}; t_k \right) ; \hat \Theta_{U_s} \right )  \right]^2,
\end{align}
where $\hat \Theta_{U_s}$ and $\hat \Theta_{R_o}$ have been obtained in Steps 1 and 2, and the form of $V_{s,1}({\Theta_s}, t)$ is derived in Appendix.\ref{subsec:A1}. 

{\em {Step 4: Identification of $\Theta_{\mathrm{Th}}$.}} Based on~\cite{Lin:JPS:2014}, a straightforward idea to determine $\Theta_{\mathrm{Th}}$ is to fit it to the measurement data of $T_\mathrm{surf}$ and/or $T_\mathrm{core}$ given the lumped thermal model in~\eqref{Lumped-Thermal-Model}. However, the idea is hard to be applied here, because $Q$ in our model is dependent on $R_{o,T}$, as shown in (\ref{Qdot Model}), and  unavailable before   $R_{o,T}$ is identified. To overcome this issue, we choose to   use prior knowledge to guide the estimation of $\Theta_{\mathrm{Th}}$. Here, we can approximate $R_\mathrm{core}$ based on the conductivity of the cell's electrode materials and jellyroll structure. Furthermore, we can infer $R_\mathrm{surf}$ and $C_\mathrm{surf}$ from the form factors and specifications, casing material (usually aluminum), and the cooling system. Finally, $C_\mathrm{core}$ can be deduced given the cell's total heat capacity. A 2C constant current full discharge profile is used to acquire the data encompassing significant temperature changes. With the measurement data, 
we can begin from the approximate values of the parameters and continually tune them until achieving sufficient fitting accuracy to finalize $\hat \Theta_{\mathrm{Th}}$.

{\em Step 5: Identification of $\Theta_e$ and $\Theta_\mathrm{Arr}$.} Sub-circuit B will have substantial effects on $U$ only at high C-rates. Therefore, we use a 3 C constant current profile   to fully discharge the cell such that large enough $U_e$ will result and present itself into the   voltage response. This then allows to identify $\Theta_e$. In the meantime,   3 C discharging will  subject the cell to important temperature increases, which, in turn, will drive down     $\Theta_\mathrm{Arr}$-dependent $R_{s,T}$ and $R_{o,T}$ and influence the voltage response.  As such, we need to consider the estimation of $\Theta_e$ and  $\Theta_\mathrm{Arr}$ together. The  following data fitting problem   can be formulated: 
\begin{align}\label{Thetae-ThetaArr-Identify}
\hat  \Theta_\mathrm{e}, \hat  \Theta_\mathrm{Arr}  = \arg \min_{ \Theta_\mathrm{e},    \Theta_\mathrm{Arr} } \sum_{t_k}  \left[ U(t_k) -  R_{o,T}\left( \hat \Theta_{R_o}, \Theta_\mathrm{Arr}, T_{t_k}; t_k \right)  I(t_k) - U_s \left (V_{s,1}\left({  \hat \Theta_s}, \Theta_\mathrm{Arr}, T_{t_k}; t_k \right) ; \hat \Theta_{U_s} \right ) - U_e\left( \Theta_e; t_k \right) \right]^2. 
\end{align}
Here, $U_e$ depends on $V_{e,1}$ and $V_{e,3}$ as shown in~\eqref{Elec-Potential}, and the explicit form of $V_{e,1}$ and $V_{e,3}$ is shown in Appendix.\ref{subsec:A2}.
Note that  no closed-form expression of $U_s$ exists in this step, as  the changing $R_{s,T}$ makes sub-circuit A   become a time-varying system. It  is thus impossible to solve the   problem in~\eqref{Thetae-ThetaArr-Identify} using nonlinear optimization. To alleviate the difficulty, we suggest to apply some empirical tuning. Specifically, we can pick a sample of $\Theta_\mathrm{Arr}$ using prior knowledge,   then estimate $\Theta_e$ by solving the above data fitting problem, and iterate this procedure until getting the lowest possible fitting errors.  Despite the time and effort needed, this iterative method is often found effective with a sufficient number of tries.

The above steps together constitute our parameter identification approach  for the BattX model. The following remarks   summarize our further insights.  

\begin{remark} \setstretch{1.5}
We  point out that the data fitting problems outlined in Steps 1-5 are non-trivial to solve, as they entail nonlinear nonconvex optimization. The nonconvexity can easily get the parameter search stuck in  local minima to produce physically meaningless parameter estimates. To mitigate the issue, it is   sensible to  constrain the search within
a believably correct parameter space~\cite{Tian:JES:2020}.  Specifically, one can   set up    approximate lower and upper bounds for every possible
parameter and then limit the numerical optimization within the resultant parameter space. The prior knowledge used to establish such bounds can be derived from both experience and observation or analysis of the measurement data. Other helpful ways to overcome the local minima issue include adding regularization terms that encode prior knowledge of some parameters and applying different initial guesses to repeatedly run the numerical optimization~\cite{Tian:JES:2020}.   
\end{remark}

\begin{remark} \setstretch{1.5}
We consider Samsung INR18650-25R cells (see Section~\ref{Sec:Validation} for the specifications) as a baseline when selecting the discharging C-rates in each step of the above approach, because they are used in the experimental validation of the BattX model (see Section~\ref{Sec:Validation}). However, a user or practitioner may need to adjust the specific C-rates, depending on the cells to apply the model to. The overall guiding rule is   the same---using current profiles of different C-rates to excite different dynamic processes to obtain  data informative for the identification of  the parameters associated with each process. 
\end{remark}

\section{Experimental Validation of the BattX Model}\label{Sec:Validation}

This section offers the experimental validation of    the proposed BattX model. All the experiments were conducted on  a Samsung INR18650-25R cell  with   NCA cathode and graphite anode using  a PEC\textsuperscript{\tiny\textregistered} SBT4050 battery tester.  The    cell's nominal capacity is 2.5 Ah,     nominal voltage is 3.6 V,     maximum cut-off voltage is 4.2 V,     minimum cut-off  voltage is 2.5 V, and    maximum continuous discharge current is 20 A. 
The tester is able to run  charging or discharging tests of up to 40 V and 50 A under arbitrary current or power load profiles.  The experiments comprised two   parts. The first part collected  datasets following the parameter identification approach in Section~\ref{sec:ParamID} to identify   the model parameters. In the second part, new datasets were generated  to evaluate the predictive capability of  the identified model. 

\subsection{Model Identification}

The experiments and model identification procedure are as follows.

\begin{figure}[t]
\centering
\includegraphics[width = 0.49\textwidth]{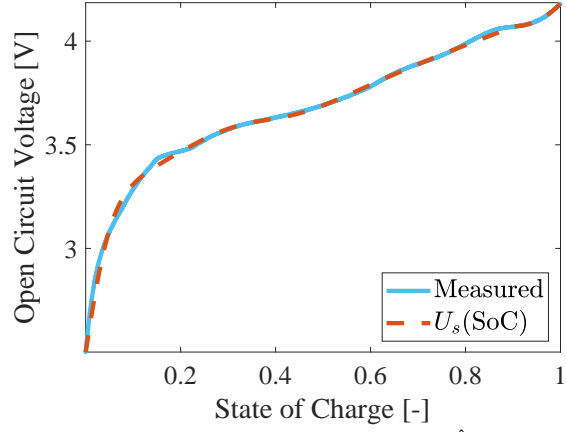}
\caption{SoC/OCV curve fitting based on $\hat \Theta_{U_s}$.}
\label{Fig:SoC-OCV-Fitting}
\end{figure}

\begin{figure}[htbp]
\centering
\begin{subfigure}{0.49\textwidth}
	\includegraphics[width=80mm]{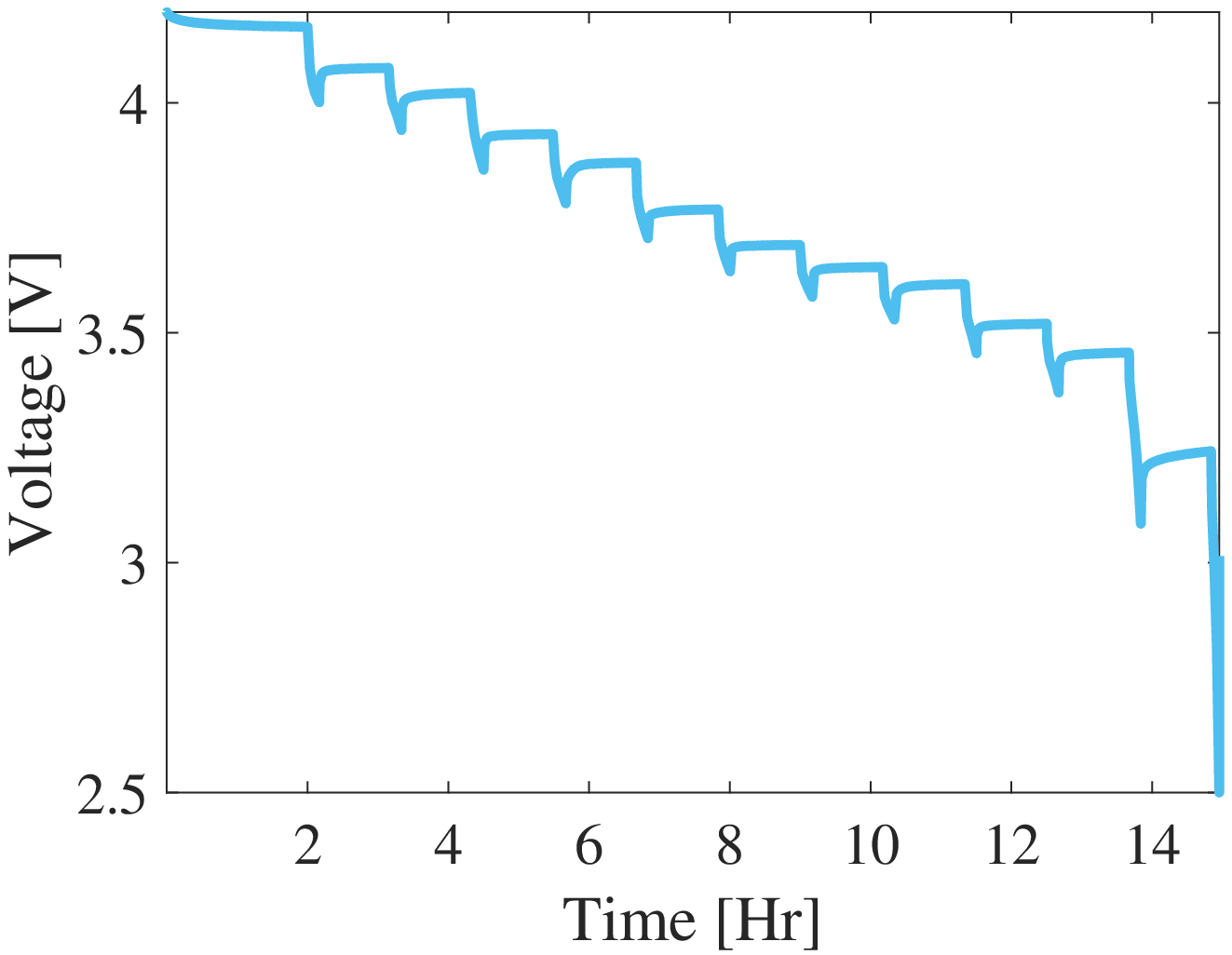}
	\caption{}
	\label{Fig:Pulse-Load-Profile}
\end{subfigure} \hfill
\begin{subfigure}{0.49\textwidth}
	\includegraphics[width=80mm]{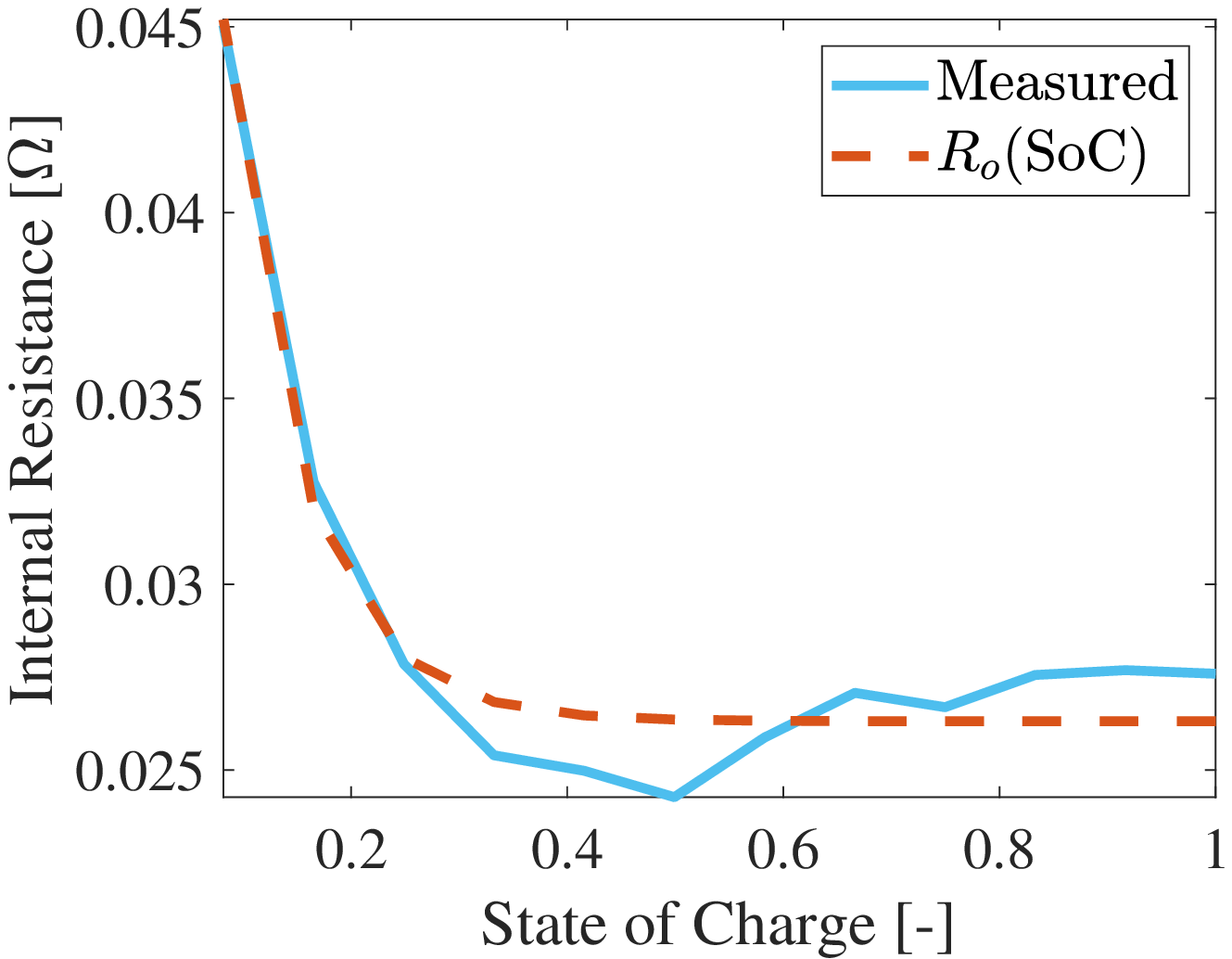}
	\caption{}
	\label{Fig:Ro-Fitting}
\end{subfigure}
\caption{Identification of $\Theta_{R_o}$: (a) terminal voltage profile under intermittent discharging at 0.5 C  to identify $\Theta_{R_o}$; (b)   fitting of $R_o\mathrm(\mathrm{SoC})$ with $\tilde R_o$ based on $\hat \Theta_{R_o}$.} 
\label{fig:Ro}
\end{figure}

\begin{figure}[t]
\centering
\includegraphics[width = 0.49\textwidth]{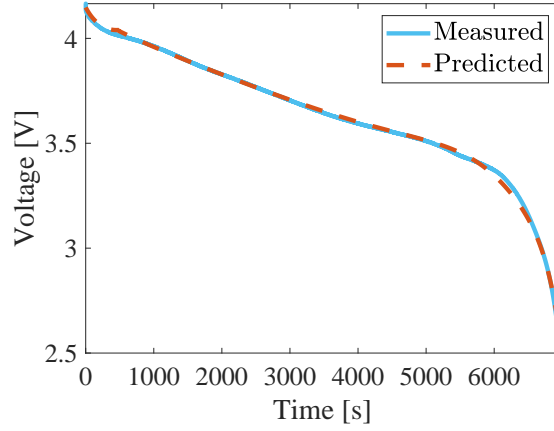}
\caption{Terminal voltage fitting under 0.5 C constant-current discharging based on $\hat \Theta_s$.}
\label{Fig:U-Fitting-under-Thetas}
\end{figure}

\begin{figure}[t]
\centering
\includegraphics[width = 0.49\textwidth]{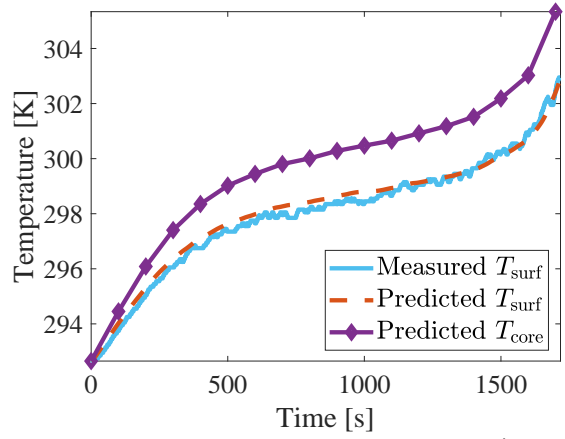}
\caption{Temperature fitting and prediction based on   $\hat \Theta_\mathrm{Th}$.}
\label{fig:Thermal-Fitting}
\end{figure}

\begin{figure}[!htbp]
\centering
\includegraphics[width = 0.49\textwidth]{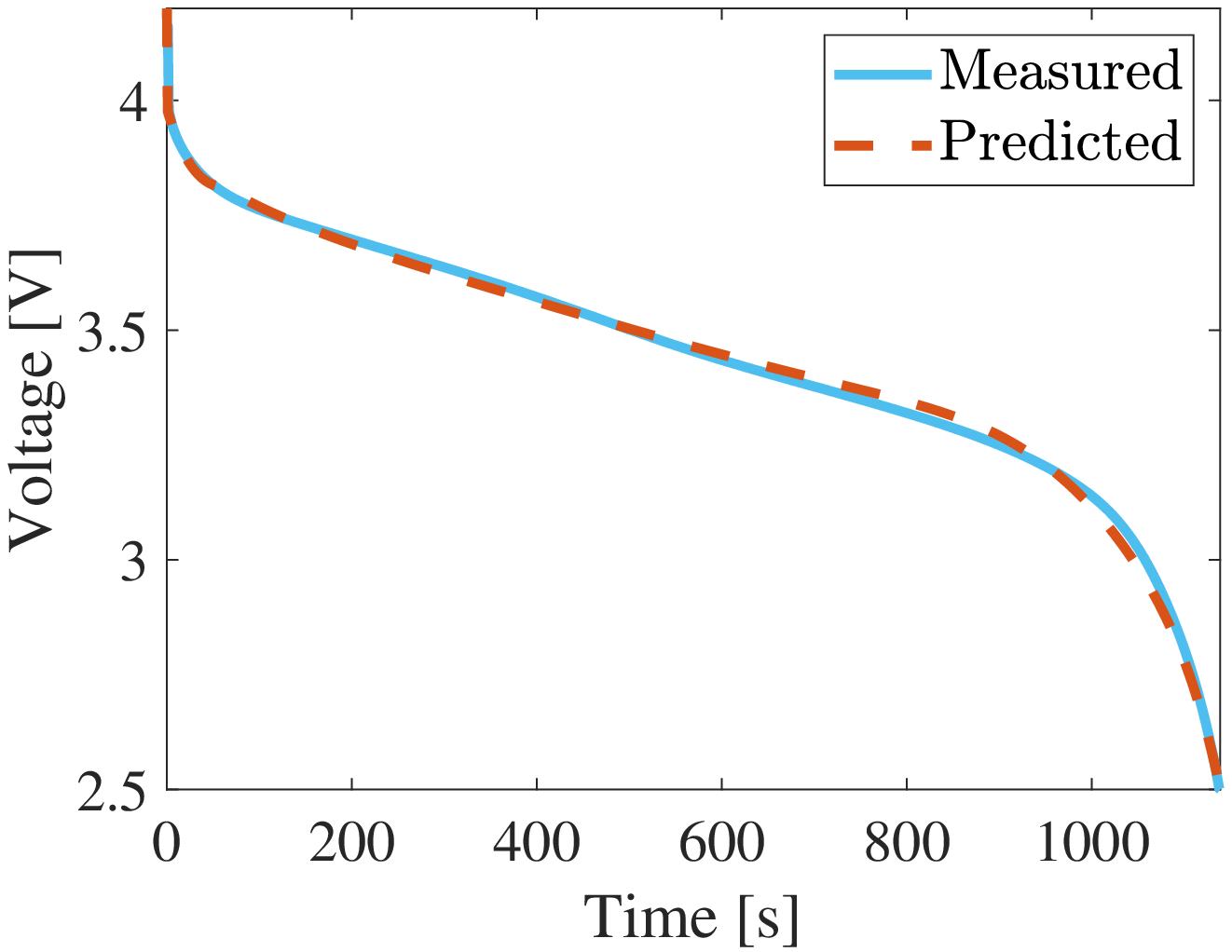}
\caption{Terminal voltage fitting under 3 C constant-current discharging based on $\hat \Theta_e$.}
\label{fig:ElectrolyteFitting}
\end{figure}

\begin{itemize}

\item Based on Section~\ref{sec:ParamID}, we first charged the cell using the popular constant-current/constant-voltage method, let it rest for one hour, and then fully discharged it using a 1/30 C constant-current load. We calculated the total capacity $2.55$ Ah using the Coulomb counting method and used the voltage data to find out $\hat \Theta_{U_s}$ based on~\eqref{Data-Fitting-Theta-Us}, which is given by
\begin{align*}
\hat \Theta_{U_s} &=  \left\{-9.048, -2.360, -12.986, 0.010, 13.036, -32.840, -0.087, 2.359, \right.\\
&\quad\quad\quad  \left. -14.863, 0.055, -0.788, -7.136, 0.966, 31.132, -3.414, 0.513, 1.816\right\}.
\end{align*}
The  SoC/OCV fitting result under the obtained $\hat \Theta_{U_s}$ is shown in Fig.~\ref{Fig:SoC-OCV-Fitting}.

\item Next, the cell was charged to full again idled for one hour, and then discharged under a 0.5 C pulse load profile. Specifically, a load was applied for five minutes, followed by a two-hour rest, and this cycle continued until the cut-off voltage was met. Fig.~\ref{fig:Ro} shows the profile, which includes a total of 12 pulses.  With the data, we  calculated $R_o$ at different SoC via~\eqref{Ro-Calculation} and then used~\eqref{Theta-Ro-Identify} to compute $\hat \Theta_{R_o}$ as   shown in Table~\ref{tab:Us-Uth}. The reconstructed $R_o$ is compared with the measurements in Fig.~\ref{fig:Ro}.

\item Going further, we fully charged the cell again as in the previous steps, and then fully discharged it using a 0.5 C constant-current load, with the objective of identifying $\Theta_s$. As explained in Section~\ref{sec:ParamID}, we could impose a pre-determined relation like~\eqref{Csi-Rsi-Relation} to reduce the number of parameters to estimate. Here, we let  the spherical particle be discretized into five finite volumes,
and the resulting $\eta_i$ and $\sigma_j$ are
\begin{align*}
\eta_i &=  \left\{1, 0.6066, 0.3115, 0.1148, 0.0164 \right\}, \\
\sigma_j &=  \left\{1, 1.77, 4.00, 15.98 \right\}.
\end{align*} 
Then,~\eqref{Thetas-Identify} was executed to determine $\hat \Theta_s$. Fig.~\ref{Fig:U-Fitting-under-Thetas} illustrates a comparison between the predicted terminal voltage (with the dynamics of sub-circuits B and C neglected) based on  $\hat \Theta_s$  and the measurements. Table \ref{tab:Us-Uth} shows the estimation for $\hat \Theta_s$.

\item Then, we ran a 2 C constant-current discharging test to collect the temperature data. The cell's surface temperature increased by about 10 K throughout the test. We leveraged prior knowledge and  empirical tuning to determine $  \Theta_\mathrm{Th}$, as suggested in Section~\ref{sec:ParamID}. While the procedure is coarse-grained, we obtained $\hat \Theta_\mathrm{Th}$ that leads to accurate fitting with the surface  temperature data and physically reasonable estimation of the core temperature, as shown in Fig.~\ref{fig:Thermal-Fitting}.  Table~\ref{tab:Us-Uth} summarizes the numerical estimates of $\hat \Theta_\mathrm{Th}$  .

\item Finally, the cell was fully discharged at a constant current of 3 C to excite the cell's  electrolyte dynamic and thermal  behavior more discernible, for the purpose of identifying $\Theta_e$ and $\Theta_\mathrm{Arr}$. Following Section~\ref{sec:ParamID}, we iteratively tuned $\hat \Theta_\mathrm{Arr}$ and then ran~\eqref{Thetae-ThetaArr-Identify} to find $\hat \Theta_e$ until the achievement of both physically sound estimates and accurate  voltage data  fitting. Fig.~\ref{fig:ElectrolyteFitting} shows that the   BattX model based on all the identified parameters fits well the measured voltage, and Table~\ref{tab:Ue-Arr} shows the estimation results.

\end{itemize}

\begin{table}
\centering
\begin{tabular}{|c|c|c|c|c|c|c|c|c|c|}
\hline
Name &  $\gamma_1$ & $\gamma_2$ & $\gamma_3$ & $C_{s,1}$ & $R_{s,1}$ & $C_{\mathrm{surf}}$ & $R_{\mathrm{surf}}$ & $C_{\mathrm{core}}$ & $R_{\mathrm{core}}$\\
\hline
Initial Guess & 1 & 1 & 1 & 4391 & 0.090 & 7 & 6 & 20 & 1\\
\hline
Lower Bound & - & - & - & 3600 & 0.054 & 3 & 3 & 5 & 0.5\\
\hline
Upper Bound & - & - & - & 5500 & 0.167 & 12 & 20 & 50 & 7\\
\hline
Final estimate & 0.026 & 0.061 & -14.36 & 4521 & 0.114 & 10 & 7 & 40 & 4\\
\hline  
\end{tabular}
\caption{Identification summary for $\Theta_{R_o}$, $\Theta_s$, and $\Theta_{\mathrm{Th}}$: initial guesses, bound limits, and final estimates.}\label{tab:Us-Uth}
\end{table}

\begin{table}[!htbp]
\centering
\begin{tabular}{|c|c|c|c|c|c|c|}
\hline
Name & $C_e$ & $R_e$ & $\beta_1$ & $\beta_2$ & $\kappa_1$ & $\kappa_2$\\
\hline
Initial Guess & 1032 & 0.028 & 0.53 & 0.31 & 15 & 22\\
\hline
Lower Bound & 500 & 0.002 & 0.42 & 0.19 & 10 & 10 \\
\hline
Upper Bound & 5000 & 0.080 & 1.00 & 0.423 & 100 & 100\\
\hline
Final estimate & 3691 & 0.007 & 0.789 & 0.317 & 30 & 70 \\
\hline  
\end{tabular}
\caption{Identification summary for $\Theta_{\mathrm{Arr}}$ and $\Theta_{e}$: initial guesses, bound limits, and final estimates.}\label{tab:Ue-Arr}
\end{table}

From above, we have come up with an explicit setup of the BattX model for the cell. Next, we will fit the model to new datasets to assess how well it predicts. 

\subsection{Model Testing and Validation}

\begin{figure}[t]
\centering
\includegraphics[width = 0.49\textwidth]{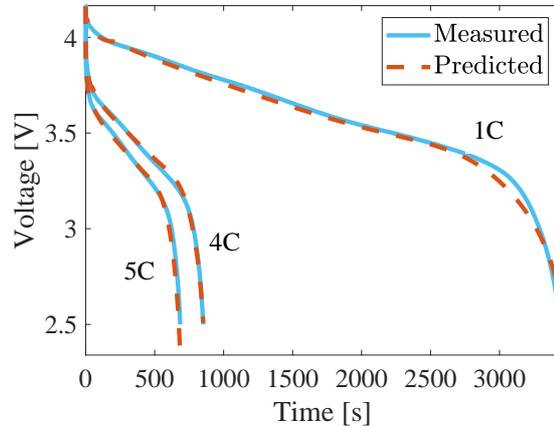}
\caption{Voltage prediction by the BattX model versus the measurements at constant-current discharging at 1, 4 and 5 C.}
\label{fig:CC_Validation}
\end{figure}

To further evaluate the obtained BattX model, we generated new datasets by applying a variety of current load profiles that span a broad range of currents. The first tests involved full discharging of the cell at a constant current  of 1 C, 4 C, and 5 C separately. Fig.~\ref{fig:CC_Validation} compares the   model's prediction of the terminal voltage prediction with the  measurement, where a close match is observed in all the three cases. Note that, even though the model was identified based on tests of only up to 3  C, it can well predict   4 C and 5 C. This suggests the model's high fidelity and interoperability. 

\begin{figure}[t]
\centering
\includegraphics[trim={0mm 15mm 0mm 15mm},clip, width = 0.9 \textwidth]{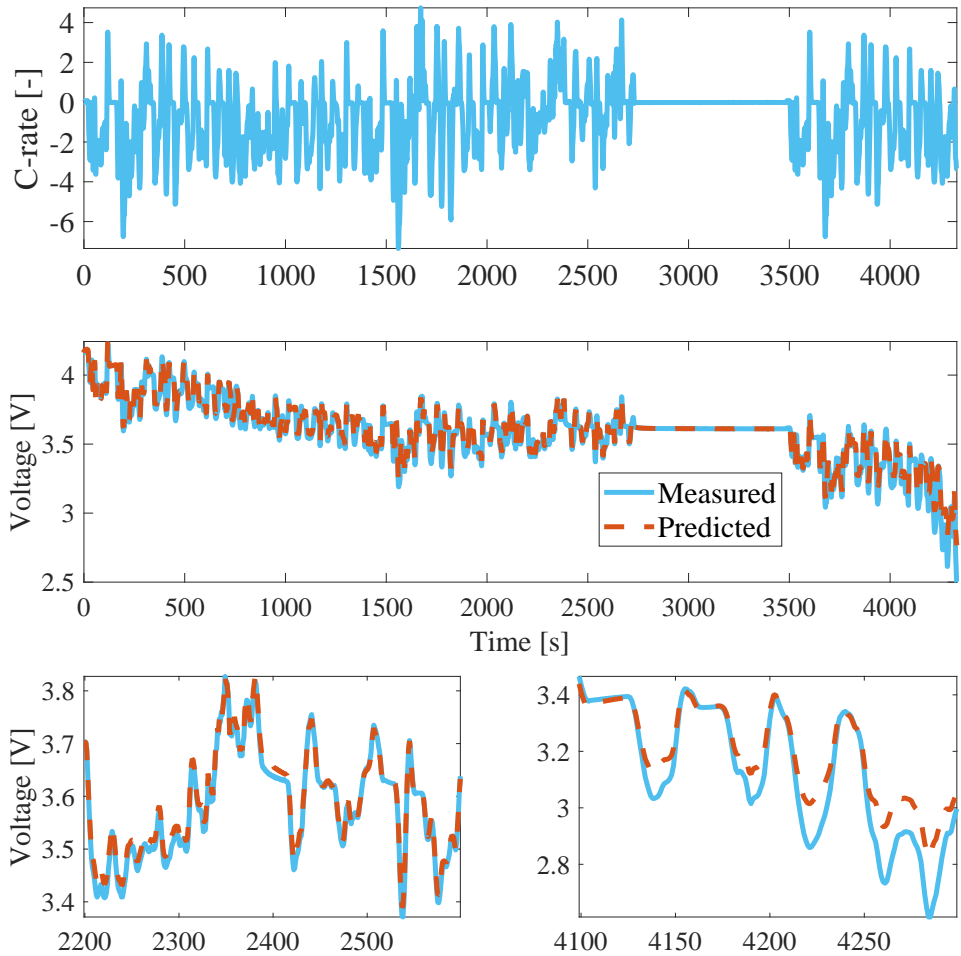}
\caption{Terminal voltage prediction by the BattX model versus the measurements in the UDDS-based test. Top: the UDDS-based current profile; middle: the voltage prediction in comparison with the measurements; bottom: magnified views within two time windows.}
\label{fig:UDDS_Validation_Volt}
\end{figure}

Further, we adopted the Urban Dynamometer Driving Schedule (UDDS) as a variable load profile and scaled it to be between $-8$ C and $5$ C. The validation of the BattX model over this dataset is shown  in Fig.~\ref{fig:UDDS_Validation_Volt}. The top figure in  Fig.~\ref{fig:UDDS_Validation_Volt} illustrates the load profile, which includes both charging and discharging as well as a rest period. The voltage prediction of  the model, as shown in Fig.~\ref{fig:UDDS_Validation_Volt}, closely follows the true voltage overall. A slight discrepancy  appears at the end of the test when the cell is about to be depleted. This is likely because the radical changes of the  internal resistance  at low SoC and high temperature are hard to be thoroughly captured. 

\begin{figure}[!htbp]
\centering
\includegraphics[ width = 0.9\textwidth]{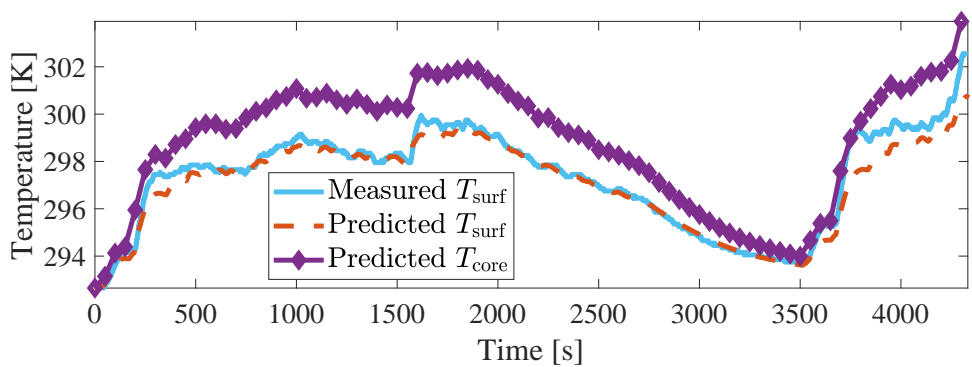}
\caption{Temperature prediction by the BattX model versus the measurements in the UDDS-based test. }
\label{fig:UDDS_Validation_Temp}
\end{figure}

 Fig.~\ref{fig:UDDS_Validation_Temp} then demonstrates the comparison of the predicted surface temperature with the measurement, showing an acceptable accuracy. The estimation of the core temperature is also given in Fig.~\ref{fig:UDDS_Validation_Temp}, which is reasonable by empirical knowledge and observation.

\begin{figure}[t]
\centering
\includegraphics[width = 0.9 \textwidth]{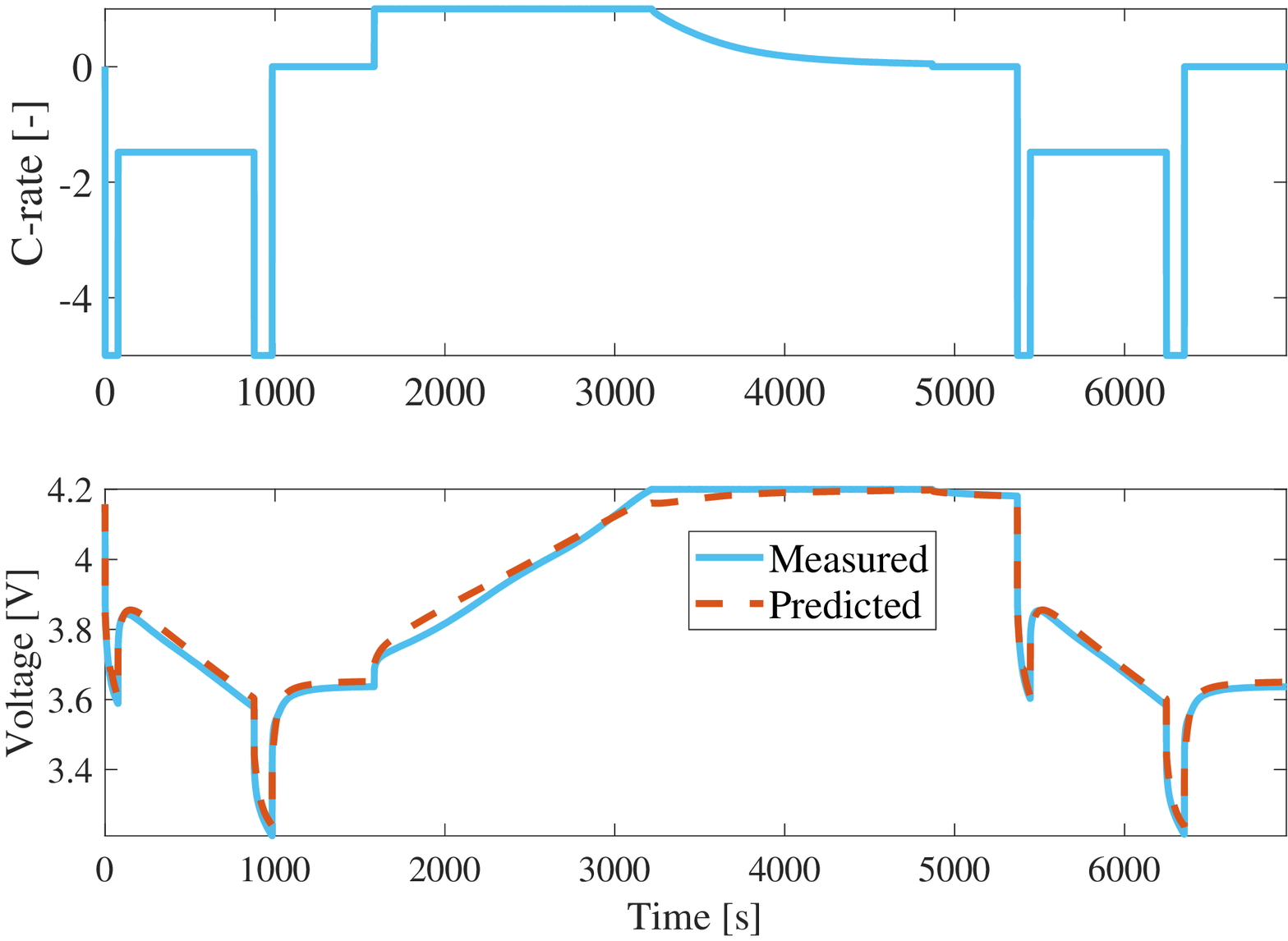}
\caption{Terminal voltage prediction by the BattX model versus the measurements in the test simulating an eVTOL operation cycle. Top: the current profile; bottom: comparison between the prediction and measurements.}
\label{fig:eVTOL_Validation_Volt}
\end{figure}

\begin{figure}[!htbp]
\centering
\includegraphics[width = 0.9 \textwidth]{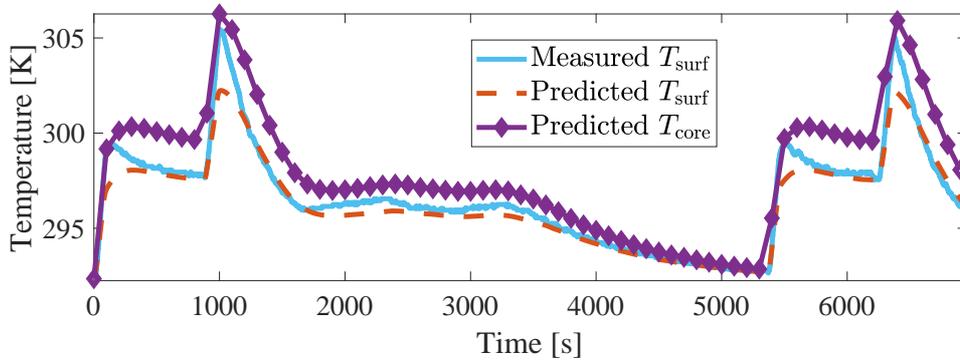}
\caption{Temperature   prediction by the BattX model versus the measurements in the test simulating an eVTOL operation cycle.}
\label{fig:eVTOL_Validation_Temp}
\end{figure}

Recently, LiB-powered eVTOL has attracted  increasing interest  as a promising solution to urban air mobility and decarbonization of aviation. A safety-critical application, eVTOL must maintain  fast and accurate monitoring   of its onboard battery system throughout a mission. 
Conventional equivalent circuit models are impossible to meet this need, as eVTOL often requires high-rate discharging---it must discharge as fast as 5 C in the takeoff and landing phases. However, the proposed BattX model holds a significant advantage to eVTOL battery performance modeling. We consider a notional eVTOL flight, which includes three phases, takeoff, cruising, and landing. The three phases involve discharging at 5 C, 1.48 C, and 5 C, respectively \cite{Sripad:arXiv:2020}. We generated a current load profile sequentially comprising a flight,   full discharge, and another flight.  Fig.~\ref{fig:eVTOL_Validation_Volt} displays the profile over time.  Fig.~\ref{fig:eVTOL_Validation_Volt} shows that the BattX model  achieves accurate prediction compared with the measurement. Especially, the accuracy is found satisfactory    at the times of high discharge rates. The surface temperature prediction in Fig.~\ref{fig:eVTOL_Validation_Temp} also well agrees with the actual temperature, and the core temperature estimation shows a realistic trend that one can trust to be  close enough with the truth.  

\section{Conclusions}
\label{Sec:Conclusion}

LiBs have found their way into many sectors as a key technology to drive forward  electrification and decarbonization. For LiB applications, computationally fast and accurate models are a bedrock for real-time monitoring and simulation to ensure their performance and safety. Although the literature has presented different dynamic models, few of them are  effective when      current loads range  from low to high. To overcome the problem, we  proposed the BattX model  in this study. This   model is an ECM  in its   form, but unlike other ECMs, it lends to interpretation as  a quasi-electrochemical model. This is  because it is designed to use separate yet coupled   circuits to approximate the lithium-ion diffusion   in the electrode and electrolyte phases, heat transfer, and nonlinear voltage behavior  in charging/discharging of a cell. With the novel design, the model  offers high predictive accuracy over broad current ranges and still retains relatively simple structures for low computational costs.  We also developed a   parameter identification approach for the model. The approach groups  the parameters  based on  the dynamic processes or components that they belong to, and then identifies the parameters of each group using experimental data. Finally,  the experimental validation showed that the BattX model has  high accuracy and fidelity across   low to high C-rates. 

\section*{Acknowledgement}

This work was supported in part by the United States National Science Foundation under Awards CMMI-1763093 and CMMI-1847651.

\section*{Appendix}
\addcontentsline{toc}{section}{Appendices}
\renewcommand{\thesubsection}{\Alph{subsection}}

\renewcommand{\theequation}{A.\arabic{equation}}
\setcounter{equation}{0}

\subsection{Derivation of $V_{s,1}$ under Constant Current $I$}
\label{subsec:A1}

In Section~\ref{sec:ParamID}, the identification of $\Theta_s$  in~\eqref{Thetas-Identify} requires the expression of $V_{s,1}$ when the applied current $I$ is constant. The derivation is as follows. 

Consider the governing equations of sub-circuit A in~\eqref{Sub-circuit-A-eqns} under the assumption in~\eqref{Csi-Rsi-Relation}, and rewrite them compactly  into the following   form:
\begin{align}\label{compact-electrode-ss}
\dot V_s(t)  &= A_s V_s(t) + B_s I (t),
\end{align}
where 
\begin{align*}
V_s &= \left[ \begin{matrix} V_{s,1} & V_{s,2} & \cdots & V_{s,N}  \end{matrix} \right]^\top,\\
A_s &= \mu_s \Omega_s, \\ 
\mu_s & = \frac{1}{C_{s,1}R_{s,1}}, \\
\Omega_s &= \begin{bmatrix}
-1 \over \eta_1 \sigma_1 & 1\over \eta_1 \sigma_1 & 0 & \cdots & \cdots & 0 \cr 
\frac{1}{\eta_2 \sigma_1} & -\frac{1}{\eta_2 \sigma_1}-\frac{1}{\eta_2 \sigma_2}  & 1 \over \eta_2 \sigma_2 & 0 & \cdots &  0 \cr
\vdots  & \vdots  & \vdots  & \ddots  & \ddots  & \vdots  \cr
0 & \cdots & \cdots & 0 &   \frac{1}{\eta_N \sigma_{N-1}} & \frac{-1}{\eta_N \sigma_{N-1}}
\end{bmatrix},\\
B_s &=  \begin{bmatrix}
     \frac{1}{C_{s,1}} & 0 & \cdots & 0  
    \end{bmatrix}^\top. 
\end{align*}
The solution to~\eqref{compact-electrode-ss} is given by
\begin{align*}
V_s(t) = e^{A_s t} V_s(0) + \int_0^t e^{A_s (t-\tau) } B_s I (\tau) d\tau.
\end{align*}
When $I$ is constant, it  becomes 
\begin{align}\label{solution-compact-electrode-ss}
V_s(t) = e^{A_s t} V_s(0) + \int_0^t e^{A_s (t-\tau) }  d\tau \cdot B_s I.
\end{align}
To find the explicit form of $V_s(t)$, we must derive the expression of $e^{A_s t}$.  
To this end, we look at $\Omega_s $ first and note that it is rank-deficient with   one zero eigenvalue. Further, assume   the other non-zero eigenvalues to be distinct, and denote the eigenvalues of $\Omega_s$ as  $\lambda_i$ for $i = 1,2,\ldots,N$ with $\lambda_1=0$.  Then, by the Cayley-Hamilton theorem, we have
\begin{align} \label{exp-As}
e^{A_st} = \left[  \Phi^{-1} \phi(\mu_s, t) \right] \otimes \Omega_s,
\end{align}
where 
\begin{align*}
\Phi &= \begin{bmatrix}
1 & \lambda_1 & \cdots & \lambda_1^{N-1}\\
1 & \lambda_2 & \cdots & \lambda_2^{N-1}\\
\vdots & \vdots & \ddots & \vdots  \\
1 & \lambda_N & \cdots & \lambda_N^{N-1}
\end{bmatrix},\\
\phi(\mu_s, t) &= \begin{bmatrix}
1 & e^{\mu_s \lambda_2 t}  & \cdots & e^{\mu_s \lambda_N t} 
\end{bmatrix}^\top.
\end{align*}
The operator $\otimes$ is defined as
\begin{align*}
a \otimes A = \sum_{i=1}^n a_i A^{i-1},
\end{align*}
for $a \in \mathbb{R}^{n\times 1}$ and $A \in \mathbb{R}^{n\times n}$.
Inserting~\eqref{exp-As} into~\eqref{solution-compact-electrode-ss}, we obtain
\begin{align}\label{Vs-expression} \nonumber
V_s(t) &=\left[ \Phi^{-1} \phi(\mu_s, t)\right]  \otimes \Omega_s \cdot V_s(0) + \int_0^t  \left[ \Phi^{-1} \phi(\mu_s, t-\tau)\right]  \otimes \Omega_s d \tau \cdot B_s I \\ \nonumber
&=\left[ \Phi^{-1} \phi(\mu_s, t)\right]  \otimes \Omega_s \cdot V_s(0) + \left[ \Phi^{-1} \int_0^t   \phi(\mu_s, t-\tau)  d \tau \right]  \otimes \Omega_s \cdot  B_s I\\
&=  \left[ \Phi^{-1} \phi(\mu_s, t)\right]  \otimes \Omega_s \cdot V_s(0)  + \left[ \Phi^{-1} \left( \bar \phi(\mu_s,t) - \bar \phi(\mu_s,0)  \right) \right]  \otimes \Omega_s \cdot  B_s  I,
\end{align}
where 
\begin{align*}
 \bar \phi(\mu_s,t) = \begin{bmatrix}
t &  { e^{\mu_s \lambda_2 t} \over \mu_s \lambda_2 }  & \cdots & e^{\mu_s \lambda_N t} \over \mu_s \lambda_N 
\end{bmatrix}^\top.
\end{align*}
Given~\eqref{Vs-expression}, $V_{s,1}$ can be expressed as 
\begin{align*}
V_{s,1}(t)  = \mathbf{e}_1^\top V_s(t),
\end{align*}
where $\mathbf{e}_1 = \begin{bmatrix}
1 & 0 & \cdots & 0
\end{bmatrix}^\top_{N \times  1}$. 

\subsection{Derivation of $V_{e,1}$ and $V_{e,3}$ under Constant Current $I$}
\label{subsec:A2}

The explicit expressions of $V_{e,1}$ and $V_{e,3}$ under a constant current $I$ are needed to represent $U_e$ for the identification of $\Theta_e$ in~\eqref{Thetae-ThetaArr-Identify}.  We can follow similar lines in Appendix.\ref{subsec:A1} to find them out. 
Let us rewrite the governing equations of sub-circuit B in~\eqref{Sub-circuit-B-eqns} compactly as
\begin{align}\label{compact-electrolyte-ss}
\dot V_e(t) = A_e V_e(t) + B_e I (t),
\end{align}
where 
\begin{align*}
V_e &=  \begin{bmatrix} V_{e,1} & V_{s,2}  & V_{e,3}  \end{bmatrix}^\top,\\
A_e  &= \mu_e \Omega_e, \\
\mu_e &= {1 \over C_e R_e }  , \\
\Omega_e &= \begin{bmatrix}
-1 & 1 & 0 \cr 1 & -2 & 1 \cr 0 & 1 & -1
\end{bmatrix}, \\
B_e &= \begin{bmatrix}
        \frac{1}{C_e} & 0 & -\frac{1}{C_e}
\end{bmatrix}^\top.
\end{align*}
The solution to~\eqref{compact-electrolyte-ss} under a constant current $I$ is  
\begin{align}\label{solution-compact-electrode-ss}
V_e(t) = e^{A_e t} V_e(0) + \int_0^t e^{A_e (t-\tau) }  d\tau \cdot B_e I.
\end{align}
The eigenvalues of $\Omega_e$ are $0, -1, -3$, respectively. By the Cayley–Hamilton theorem, it follows that
\begin{align} \label{exp-Ae}
e^{A_et} = \left[  \Psi^{-1} \psi(\mu_e, t) \right] \otimes \Omega_e,
\end{align}
where 
\begin{align*}
\Psi &= \begin{bmatrix}
1 & 0 &  0\cr
1 & -1 & 1\cr
1 & -3  & 9
\end{bmatrix},\\
\psi(\mu_e, t) &= \begin{bmatrix}
1 & e^{-\mu_e t}   & e^{-3 \mu_e t} 
\end{bmatrix}^\top.
\end{align*}
Based on~\eqref{exp-Ae}, we can   derive that
\begin{align}\label{Ve-expression}  
V_e(t)  
&=  \left[ \Psi^{-1} \psi(\mu_e, t)\right]  \otimes \Omega_e \cdot V_e(0)  + \left[ \Psi^{-1} \left( \bar \psi(\mu_e,t) - \bar \psi(\mu_e,0)  \right) \right]  \otimes \Omega_e \cdot  B_e  I,
\end{align}
where
\begin{align*}
 \bar \psi(\mu_e,t) = \begin{bmatrix}
t &  -{e^{-\mu_e  t} \over \mu_e}   &  - {e^{- 3\mu_e   t} \over 3 \mu_e }
\end{bmatrix}^\top.
\end{align*}
With~\eqref{Ve-expression}, one can   extract $V_{e,1}$ and $V_{e,3}$  from $V_e$.

\bibliography{BattX_arXiV}

\end{document}